\documentclass[sigconf]{acmart}

\usepackage{booktabs} 
\usepackage{multirow}
\usepackage{sistyle}
\SIthousandsep{,}
\usepackage{bbm}
\usepackage[british]{babel}
\usepackage{pgfplots}
\usetikzlibrary{patterns}
\usepackage{filecontents}
\pgfplotsset{compat=1.3}
\newcommand{\specialcell}[2][c]{\begin{tabular}[#1]{@{}c@{}}#2\end{tabular}}
\definecolor{green1}{RGB}{204,203,102}
\definecolor{blue1}{RGB}{102,204,182}
\definecolor{red1}{RGB}{102,163,204}
\definecolor{purple1}{RGB}{102,33,102}
\definecolor{green2}{RGB}{255,153,102}
\definecolor{blue2}{RGB}{51,204,204}
\definecolor{red2}{RGB}{143,153,204}

\definecolor{tuatara}{RGB}{67, 67, 67}
\definecolor{aluminum}{RGB}{153,153,153}
\definecolor{silver}{RGB}{191,191,191}
\definecolor{platinum}{RGB}{228,227,228}
\definecolor{mercury}{RGB}{240,240,240}
\definecolor{gallery}{RGB}{250,250,250}
\definecolor{free_speech_aquamarine}{RGB}{0, 156, 114}
\definecolor{sun_shade}{RGB}{255, 144, 68}
\definecolor{fern}{RGB}{101,197,117}
\definecolor{french_blue}{RGB}{0, 112, 182}
\definecolor{sushi}{RGB}{117, 168, 47}
\definecolor{shakespeare}{RGB}{35, 184, 223}
\definecolor{egg_shell}{RGB}{238, 234, 215}
\definecolor{carnation}{RGB}{245, 80, 86}
\definecolor{flamingo}{RGB}{237, 88, 85}
\definecolor{jet_stream}{RGB}{188, 214, 210}
\definecolor{jelly_bean}{RGB}{45, 126, 150}
\definecolor{tree_poppy}{RGB}{246, 154, 27}

\begin{document}

\copyrightyear{2018} 
\acmYear{2018} 
\setcopyright{acmcopyright}
\acmConference[SIGIR '18]{The 41st International ACM SIGIR Conference on Research \& Development in Information Retrieval}{July 8--12, 2018}{Ann Arbor, MI, USA}
\acmBooktitle{SIGIR '18: The 41st International ACM SIGIR Conference on Research \& Development in Information Retrieval, July 8--12, 2018, Ann Arbor, MI, USA}
\acmPrice{15.00}
\acmDOI{10.1145/3209978.3209980}
\acmISBN{978-1-4503-5657-2/18/07}

\title{Modeling Diverse Relevance Patterns in Ad-hoc Retrieval}

\author{Yixing Fan$^{\dagger, \ddag}$, Jiafeng Guo$^{\dagger, \ddag}$, Yanyan Lan$^{\dagger, \ddag}$, Jun Xu$^{\dagger, \ddag}$, Chengxiang Zhai$^{\ast}$ and Xueqi Cheng$^{\dagger, \ddag}$}
\affiliation{
  \institution{${\dagger}$University of Chinese Academy of Sciences, Beijing, China\\
  $^{\ddag}$CAS Key Lab of Network Data Science and Technology, Institute of Computing Technology,\\ Chinese Academy of Sciences, Beijing, China\\
  	$^{\ast}$Department of Computer Science University of Illinois at Urbana-Champaign, IL, USA
  } 
}
\email{fanyixing@software.ict.ac.cn, {guojiafeng, lanyanyan, junxu, cxq}@ict.ac.cn, czhai@illinois.edu}

\renewcommand{\shortauthors}{F. Yixing et al.}

\begin{abstract}

Assessing relevance between a query and a document is challenging in ad-hoc retrieval due to its diverse patterns, i.e., a document could be relevant to a query as a whole or partially as long as it provides sufficient information for users' need. Such diverse relevance patterns require an ideal retrieval model to be able to assess relevance in the right granularity adaptively. Unfortunately, most existing retrieval models compute relevance at a single granularity, either document-wide or passage-level, or use fixed combination strategy, restricting their ability in capturing diverse relevance patterns. In this work, we propose a data-driven method to allow relevance signals at different granularities to compete with each other for final relevance assessment. Specifically, we propose a HIerarchical Neural maTching model (HiNT) which consists of two stacked components, namely local matching layer and global decision layer. The local matching layer focuses on producing a set of local relevance signals by modeling the semantic matching between a query and each passage of a document. The global decision layer accumulates local signals into different granularities and allows them to compete with each other to decide the final relevance score. Experimental results demonstrate that our HiNT model outperforms existing state-of-the-art retrieval models significantly on benchmark ad-hoc retrieval datasets.

\end{abstract}

%
%




\maketitle

\section{Introduction}
A central question in ad-hoc retrieval is how to learn a generalizable function that can well assess  relevance between a query and a document. One of the major difficulties for relevance assessment lies in that there might be diverse relevance patterns between a query and a document. As revealed by the evaluation policy in TREC ad-hoc task \cite{clarke2005trec,sanderson2010test}, ``a document is judged relevant if any piece of it is relevant (regardless of how small the piece is in relation to the rest of the document)\footnote{http://trec.nist.gov/data/reljudge\_eng.html}''. In other words, a document could be relevant to a query as a whole or partially as long as it provides sufficient information for users' needs. 
Such diverse relevance patterns might be highly related to the heterogeneity of long documents in ad-hoc retrieval. As discussed by Robertson and Walker \cite{robertson1994some}, there are two underlying hypotheses concerning document structures in relevance judgement, i.e. verbosity hypothesis and scope hypothesis \cite{robertson1994some}. 
With the \textit{Verbosity Hypothesis} a long document might be relevant to a query as a whole, while with the \textit{Scope Hypothesis} the relevant parts could be in any position of a long document, and thus it could be partially relevant to a query.
%
%

The diverse relevance patterns call for a retrieval model to be able to assess relevance at the right granularity adaptively in ad-hoc retrieval. Unfortunately, most existing retrieval models operate at a single granularity, either document-wide or passage-level. Specifically, document-wide approaches compare a document as a whole to a query. For example, most probabilistic retrieval methods (e.g., BM25 or language models) and learning-to-rank models \cite{freund2003efficient, joachims2006training} rely on document-wide feature statistics for relevance computation. Obviously, such document-wide approaches are difficult to model finer-granularity relevance signals, leading to potential biases on the competition between long and short documents \cite{na2015two}. On the other hand, Passage-level approaches segment a document into passages and aggregate passage-level signals for relevance computation \cite{callan1994passage, Liu2002Passage, salton1993approaches}. However, the performance of existing passage-based approaches is mixed when applied to a variety of test beds \cite{wang2008discriminative} by only using simple manually designed operations over the passage-level signals. 
There have been a few efforts \cite{bendersky2008utilizing, wang2008discriminative} trying to combine both document-wide and passage-level methods. For example, Bendersky et al. \cite{bendersky2008utilizing} integrated the query-similarity on a document and its passages using document-homogeneity. Wang et al. \cite{wang2008discriminative} combined the document retrieval results with passage retrieval results using a heuristic function~\cite{lee1997analyses}. However, by using a fixed combination strategy, these models cannot fully capture the diverse relevance patterns for different query-document pairs.

Recently, deep neural models have been applied to ad-hoc retrieval. These data-driven methods have shown their expressive power in end-to-end learning relevance matching patterns between queries and documents \cite{huang2013learning, guo2016deep, mitra2017neural}. However, most existing neural matching models, either representation-focused \cite{huang2013learning} or interaction-focused \cite{guo2016deep}, belong to the document-wide approaches. For example, the representation-focused models aim to learn a document representation to compare with the query representation, while the interaction-focused models learn from a matching matrix/histogram between a document and a query. To the best of our knowledge, so far there have been no neural matching model proposed to learn relevance signals from both document-wide and passage-level explicitly for modeling diverse relevance patterns in ad-hoc retrieval.

In this paper, we propose a data-driven method to automatically learn relevance signals at different granularities (i.e. passage-level and document-wide), and allow them to compete with each other for final relevance assessment. Specifically,
we propose a HIerarchical Neural maTching model (HiNT) which consists of two stacked components, namely local matching layer and global decision layer.
 The local matching layer focuses on producing a set of local relevance signals between a query and each passage of a document. Many well-known information retrieval (IR) heuristics that characterize the relevance matching between a query and a passage can be encoded in this layer for high quality signal generation. Specifically, we employ a spatial GRU model \cite{wan2016match} for the relevance matching between a query and a passage, which can well capture semantic relations, proximities, and term importance.
 The global decision layer aims to accumulate passage-level signals into different granularities and allow them to compete with each other to form the final relevance score.
 Flexible strategies are applied in this layer to model diverse relevance patterns.
 Specifically, we utilize a hybrid network architecture to accumulate local signals, and select signals from both passage-level and document-wide to generate the final relevance score.

We evaluate the effectiveness of the proposed model based on two representative ad-hoc retrieval benchmark datasets from the LETOR collection \cite{qin2010letor}. For comparison, we take into account several well-known traditional retrieval models, learning to rank models, and deep neural matching models. These models belong to document-wide, passage-level and hybrid approaches. The empirical results show that our model can outperform all the baselines in terms of all the evaluation metrics.
We also provide detailed analysis on HiNT model, and conducte case studies to verify the diverse relevance patterns captured by our model over different query-document pairs.


\begin{figure*}[!tbp]
\centering
\includegraphics[scale=0.45]{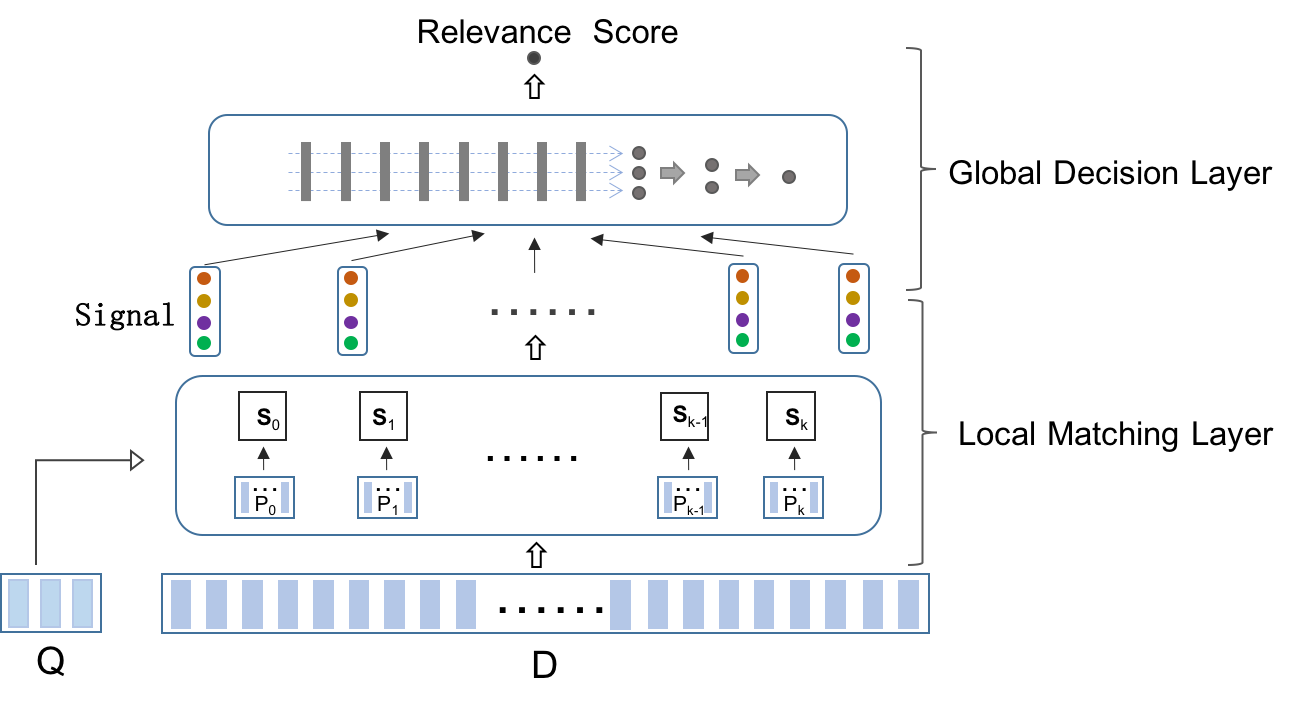}
\caption{The Architecture of the Hierarchical Neural Matching Model.}
\label{fig:model}
\end{figure*}

\section{Related Work}
A large number of retrieval methods have been proposed in the past few decades \cite{callan1994passage, robertson1994some, qin2010letor, mitra2017neural}. Without loss of generality, we divide existing methods into three folds, namely document-wide approaches, passage-level approaches and hybrid approaches, based on what kind of relevance signals they rely on for relevance assessment. We will briefly review these studies in the follows.

\subsection{Document-wide Approaches}
Document-wide approaches, by its name, collect and make relevance assessment based on document-wide signals. There have been a large number of retrieval models under this branch. Firstly, traditional retrieval models \cite{amati2002probabilistic,robertson1994some, zhai2001study}, collect lexical matching signals (e.g., term frequency) from the whole document, and make relevance assessment under some probabilistic framework. For example, the well-known BM25 model \cite{robertson1994some} collects term frequencies and document length and employs a scoring function derived under the 2-poisson model to compute relevance based on these document-wide signals. 
Secondly, most machine learning based retrieval methods, including learning to rank models and deep learning models, are also belong to this branch. For learning to rank methods~\cite{burges2010ranknet,joachims2006training, xu2007adarank}, they typically involve two stages, a feature construction stage and a model learning stage. The feature construction stage could be viewed as to define (or collect) relevance signals for a document given a query. Typically, there are three type of features, including query dependent features, document independent features, and query-document dependent features. Most of these features are defined at document level, such as tf-idf scores, BM25 scores, and PageRank. Based on these features, linear \cite{joachims2006training} or non-linear \cite{burges2010ranknet,xu2007adarank} models are learned to produce the relevance score by optimizing some ranking based loss functions. 
For deep learning methods \cite{huang2013learning, guo2016deep, pang2016study,mitra2017learning}, they can be categorized into two types according to their architectures~\cite{guo2016deep}, namely representation-focused models and interaction-focused models. The representation-focused models, such as ARCI~\cite{hu2014convolutional} and DSSM~\cite{huang2013learning}, aim to learn a low-dimensional abstractive representation for the whole document, and compare it with the query representation. The interaction-focused models, such as DRMM~\cite{guo2016deep} and Duet~\cite{mitra2017learning}, learn from a matching matrix/histogram between a document and a query to produce a set of document-wide matching signals for final relevance prediction. Although in interaction-focused models, document-wide signals are usually generated from local signals, there is no competition between document-wide signals and local signals for capturing diverse relevance patterns.

Since document-wide approaches take document as a whole, these methods are often difficult to model fine-granularity relevance signals. Meanwhile, by using document-wide statistics as relevance signals, it often leads to certain bias on the competition between long and short documents \cite{na2015two}, since long documents are likely to contain stronger signals on average.

\subsection{Passage-level Approaches}
As opposed to document-wide methods, passage-level methods collect signals from passages to make relevance assessment on the document. Note here we focus on document retrieval using passage-level signals, and will not include the work taking passages as retrieval units \cite{kaszkiel1997passage}.

In passage-level methods, documents are usually pre-segmented into small passages. Callan \cite{callan1994passage} studied how passages can be defined, and how passage signals can be incorporated into document retrieval. In \cite{Liu2002Passage}, Liu et al.~computed a language model for each passage as relevance signals. The final assessment is made by choosing the highest score from all the passages. Their results showed passage-based retrieval can provide more reliable performance than full document retrieval.
In \cite{Lv2009Positional}, Lv et al.~proposed a positional language model in which the relevance score at each position can propagate to nearby positions within certain distance. In other words, each position can be viewed as a "soft passage" by aggregating language model scores within a context window. Based on the signals at each position, they proposed three strategies, namely \emph{best position strategy}, \emph{multi-position strategy} and \emph{multi-$\sigma$ strategy}, for final relevance assessment.

As we can see, passages are convenient text units for local signal collection to support flexible relevance assessment over documents. However, previous passage-level methods often employed simplified aggregation strategies, thus cannot well capture the diverse relevance patterns for different query-document pairs.

\subsection{Hybrid Approaches}

There also have been a few efforts \cite{bendersky2008utilizing, wang2008discriminative, xi2001incorporating} trying to combine both document-wide and passage-based methods. For example, Callan \cite{callan1994passage} conducted experiments on four TREC 1 and 2 collections and concluded that it was always better to combine document-wide scores and passage-level scores. A later study by Xi et al. \cite{xi2001incorporating} re-examined fixed-size window passages on TREC 4 and 5. Contrary to Callan \cite{callan1994passage}, they did not obtain an improvement by linearly combine passage-level score and document-level score.  
Wang et al.~\cite{wang2008discriminative} proposed a discriminative probabilistic model in capturing passage-level signals, and combined the document-level scores and passage-level scores through a heuristic function (i.e., CombMNZ function \cite{lee1997analyses}). However, by using a unified combination strategy, these models cannot fully capture the diverse relevance patterns in different query-document pairs, leading to the mixed performance on different datasets \cite{xi2001incorporating, wang2008discriminative}.

\section{Hierarchical neural matching model}
In this work, we introduce a HIerarchical Neural maTching (HiNT) model for ad-hoc retrieval to explicitly model the diverse relevance patterns. In an abstract level, the model consists of two stacked components, namely local matching layer and global decision layer. The local matching layer employs deep matching networks to automatically learn the passage-level relevance signals ; 
The global decision layer accumulates passage-level signals into different granularities and allows them to compete with each other to generate the final relevance score. The architecture is illustrated in Figure \ref{fig:model}. We will describe these components in detail in the follows.

\subsection{Local Matching Layer}
The local matching layer focuses on producing a set of passage-level relevance signals by modeling the relevance matching between a query and each passage of a document.
Formally, each document $\mathbf{D}$ is first represented as a set of passages $\mathbf{D}=[P_1, P_2, ..., P_K]$, where $K$ denotes the number of passages in a document. Then, a set of passage-level relevance signals $\mathbf{E}=[\mathbf{e}_1, \mathbf{e}_2, ..., \mathbf{e}_K]$ are produced by applying some relevance matching model $f$ over a query $Q$ and each passage $P_i$.
	$$e_i = f(P_i, Q),\!\!\quad\quad i\!=1,\dots,K.\nonumber$$
There are two major questions concerning this layer, i.e., how to define the passage $P_i$ and how to define the relevance matching model $f$. For passages, there have been three types of definitions: discourse, semantic, and window \cite{callan1994passage}. Discourse passages are defined based upon textual discourse units (e.g., sentences, paragraphs, and sections). Semantic passages are based upon the subject or content of the text (e.g., TextTiling). Window passages are obtained based upon a number of words. Among these methods, window passage is the most widely adopted due to its simplicity but surprisingly effectiveness as demonstrated by many previous passage-level retrieval models \cite{bendersky2008utilizing,callan1994passage, wang2008discriminative,salton1993approaches}.

For the relevance matching model, in general, any model that can address the relevance matching between a query and a passage can be leveraged here. For example, one may employ statistical language model \cite{zhai2001study}, or use manually defined features \cite{qin2010letor,wu2007retrospective}, or even employ some deep models for text matching \cite{huang2013learning, wan2016match}. However, the quality of the passage-level signals produced by the relevance matching model is a critical foundation for the final relevance assessment. Therefore, we argue that a relevance matching model should be able to encode many well-known IR heuristics to be qualified in this layer. According to previous studies, such heuristics at least include the modeling of exact matching and semantic matching \cite{fang2006semantic, guo2016deep}, proximity \cite{tao2007exploration}, term importance~\cite{guo2016deep} and so on.

Based on these above ideas, in this work, we propose to use fixed-size window to define the passage, and employ an existing spatial GRU model \cite{wan2016match} for the relevance matching between a query and each passage. 
This spatial GRU model is good at modeling the matching between two pieces of texts based on primitive word features, and has shown better performances as compared with other deep matching models \cite{wan2016match}. We now describe the specific implementation in the follows.

\subsubsection{The Input Layer}
Following the idea in \cite{guo2016semantic}, term vectors are employed as basic representations so that rich semantic relations between query and document terms can be captured.
Formally, both query and document are represented as a sequence of term vectors denoted by $ \mathbf{Q}={[\mathbf{w}_1^{(Q)},..., \mathbf{w}_M^{(Q)}]}$ and $\mathbf{D}={[\mathbf{w}_1^{(D)},...,\mathbf{w}_N^{(D)}]}$, where $\mathbf{w}_i^{(Q)}, i=1,...,M$ and $\mathbf{w}_j^{(D)},j=1,...,N$ denotes a query term vector and a document term vector, respectively.
To obtain the passages, we follow previous approaches \cite{callan1994passage,bendersky2008utilizing,wang2008discriminative} to use fixed-size sliding window to segment the document into passages. In this way, the passage is defined as $\mathbf{P}=[\mathbf{w}_1^{(P)},...,\mathbf{w}_{L}^{(P)}]$, where $L$ denotes the window size.

\subsubsection{Deep Relevance Matching Network}

\begin{figure}[!tbp]
\centering
\includegraphics[scale=0.45]{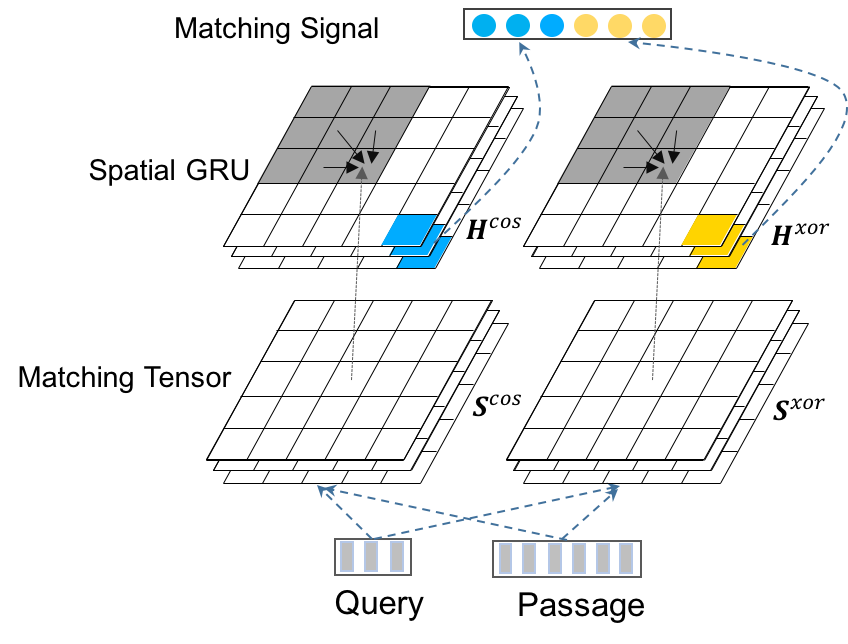}
\caption{The Architecture of Relevance Matching Network.}
\label{fig:evidence}
\end{figure}

\begin{figure*}[!tbp]
\centering
\includegraphics[scale=0.5]{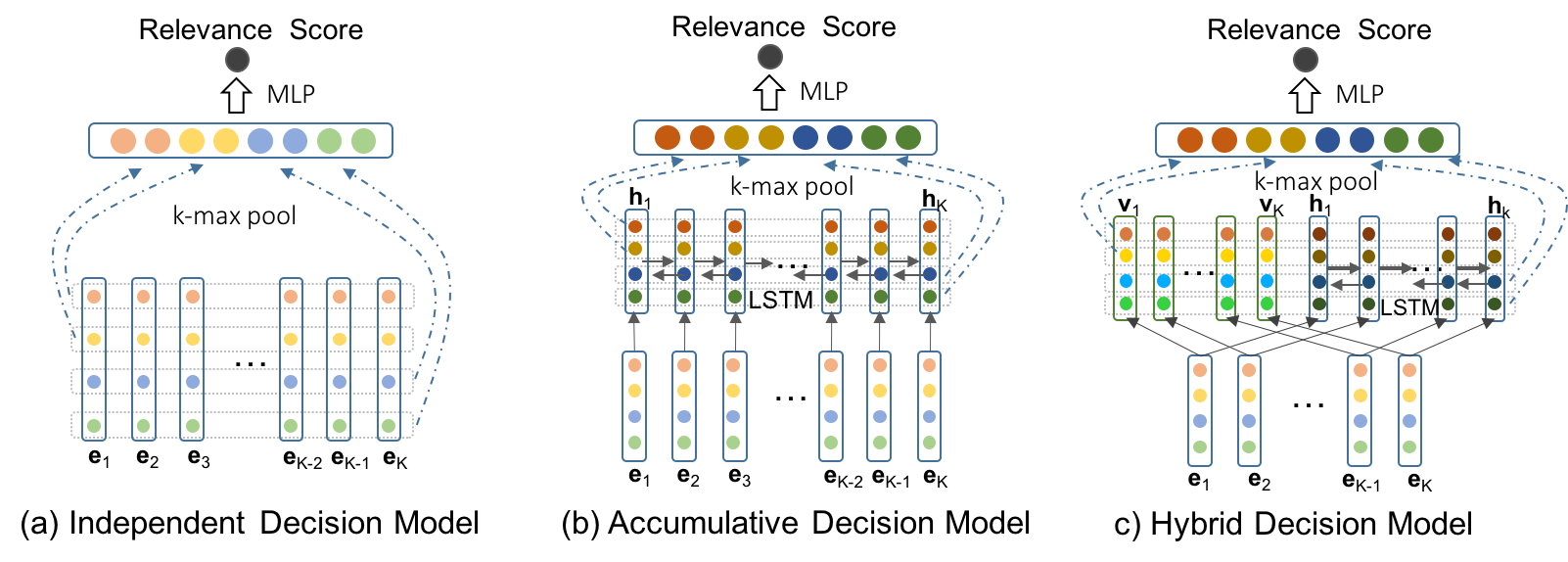}
\caption{Different decision models based on the collected passage signals.}
\label{fig:decisions}
\end{figure*}

The architecture of the relevance matching network is shown in Figure \ref{fig:evidence}.
Firstly, the term-level interaction matrix is constructed based on the term vectors from the query-passage pair. Here we constructed two matching matrices, a semantic matching matrix $\mathbf{M}^{\cos}$ and an exact matching matrix (i.e., xor-matrix) $\mathbf{M}^{\mathit{xor}}$, defined as follows:
$$\mathbf{M}^{\cos}_{ij} = \frac{\mathbf{w}_i^{(Q)} \mathbf{w}_j^{(P)}}{|\mathbf{w}_i^{(Q)}|\cdot|\mathbf{w}_j^{(P)}|},$$
$$\mathbf{M}^{\mathit{xor}}_{ij} = \begin{cases}
	1,\quad if\, \mathbf{w}_i^{(Q)} = \mathbf{w}_j^{(P)} \\ 0,\quad  otherwise  \end{cases}.$$
The key idea of two input matrices is to distinguish the exact matching signals from the semantic matching signals explicitly since the former provides critical information for ad-hoc retrieval as suggested by \cite{fang2006semantic,guo2016deep}. Note that in $\mathbf{M}^{\cos}$ exact matching and semantic matching signals are mixed together.
To further incorporate term importance, we extend each element of $\mathbf{M}_{ij}$ to a three-dimensional vector $\mathbf{S}_{ij} = [\mathbf{x}_i, \mathbf{y}_j, \mathbf{M}_{ij}]$ by concatenating two corresponding compressed term embeddings as in \cite{pang2017deeprank}, where $\mathbf{x}_i = \mathbf{w}_i^{Q} * \mathbf{W}_s$ and $\mathbf{y}_j = \mathbf{w}_j^{(P)} * \mathbf{W}_s$, here, $\mathbf{W}_s$ is the transformation parameter to be learned.

Based on these two query-passage interaction tensors, a spatial GRU (Gated Recurrent Units) is applied to generate the relevance matching evidences. The spatial GRU, also referred to as 2-dimensional Gated-RNN, is a special case of multidimensional RNN \cite{graves2009offline}. It is a recursive model which scans the input tensor from top left to bottom right:
	 $$\overrightarrow{\mathbf{H}}^{\cos}_{ij} = g(\overrightarrow{\mathbf{H}}^{\cos}_{i-1,j}, \overrightarrow{\mathbf{H}}^{\cos}_{i,j-1}, \overrightarrow{\mathbf{H}}^{\cos}_{i-1,j-1}, \mathbf{S}^{\cos}_{ij}),$$
	 $$\overrightarrow{\mathbf{H}}^{\mathit{xor}}_{ij} = g(\overrightarrow{\mathbf{H}}^{\mathit{xor}}_{i-1,j}, \overrightarrow{\mathbf{H}}^{\mathit{xor}}_{i,j-1}, \overrightarrow{\mathbf{H}}^{\mathit{xor}}_{i-1,j-1}, \mathbf{S}^{\mathit{xor}}_{ij}),$$
where $g$ denotes the spatial GRU unit as described in \cite{wan2016match}, $\overrightarrow{\mathbf{H}}^{\cos}_{ij}$ and $\overrightarrow{\mathbf{H}}^{\mathit{xor}}_{ij}$ denotes the hidden state of the spatial GRU over $\mathbf{S}^{\cos}$ and $\mathbf{S}^{\mathit{xor}}$, respectively. We can take the last hidden representation $\mathbf{H}_{M,L}$ as the matching output. The local relevance evidence $\overrightarrow{\mathbf{e}}$ is then generated by concatenating the two matching outputs:
	 $$\overrightarrow{\mathbf{e}} = [\overrightarrow{\mathbf{H}}^{\cos}_{M,L}, \overrightarrow{\mathbf{H}}^{\mathit{xor}}_{M,L}].$$
	
Furthermore, in order to enrich the relevance signals, we also applied the spatial GRU in the reverse direction, i.e., from bottom right to top left. The final passage-level signal is defined as the concatenation of the two-direction matching signals:
$$\mathbf{e} = [ \overrightarrow{\mathbf{e}},\, \overleftarrow{\mathbf{e}} ].$$

\subsection{Global Decision Layer}
Based on passage-level signals generated in the previous step, the global decision layer attempts to accumulate these signals into different granularities and allow them to compete with each other for final relevance assessment. As we have discussed before,
the relevance patterns of a query-document pair can be rather flexible and diverse, allowing a document to be relevant to a query partially or as a whole.
Accordingly, the global decision layer is expected to be able to accommodate various decision strategies, rather than using some restrictive combination rules \cite{bendersky2008utilizing, wang2008discriminative} .

In this work, we propose to employ a hybrid neural network architecture which has sufficient expressive power to support flexible relevance patterns. Before we describe the hybrid model, we first introduce two basic models under some simplified relevance assumptions.

\begin{description}
\item{1.} \textbf{Independent Decision (ID) Model} assumes the independence among passage-level signals, and selects top-k signals directly for final relevance assessment. This model is under the assumption that a document is relevant if any piece of it can provide sufficient relevance information. Specifically, as shown in Figure \ref{fig:decisions}(a), a dimension-wise k-max pooling layer is first applied over the passage-level signals to select top-k signals, and the selected signals are then concatenated and projected into a multi-layer perceptron to get the final decision score.
	
\item{2.} \textbf{Accumulative decision (AD) Model} accumulates the passage-level signals in a sequential manner, and selects top-k accumulated signals for relevance assessment.
Here, we adopt long-short term memory network (LSTM) \cite{hochreiter1997long}, a powerful model for variable-length sequential data, to accumulate the relevance signals from each passage.
Specifically, as shown in Figure \ref{fig:decisions}(b), we first feed the passage-level signals into LSTM sequentially to generate the accumulated relevance signals at different positions.
Based on the accumulated relevance signals, we then apply a dimension-wise k-max pooling layer to select top-k signals, and feed the selected signals into a multi-layer perceptron for final relevance assessment. Here, we also applied the LSTM in the reverse direction to accumulate the relevance signals, as user's reading can be in any direction of the document\cite{rayner1998eye}.
	
Note here if we directly use the last/first hidden state in LSTM as signals for relevance assessment, it would reduce to a document-wide method. By using k-max pooling over all the positions, we actually assume the relevance could be based on a text span flexible in scale, ranging from multiple passages to the whole document.
\end{description}

Based on the two basic models, now we introduce the \textbf{Hybrid Decision (HD) Model} as a specific implementation of the global decision layer. The HD model is a mixture of the previous two models, and picks top-k signals from passage-level or accumulated signals for final relevance assessment. Obviously, this is the most flexible relevance model, which allows a document to be assessed as a whole or partially adaptively.
Specifically, as depicted in figure \ref{fig:decisions}(c), we allow the relevance signals from different passages to  compete with accumulated signals. Note here in order to make a fair competition, for the passage-level signals, we conduct an additional non-linear transformation to ensure a similar scale to the accumulated relevance signals.
		$$\mathbf{v}_{t} = \tanh(\mathbf{W}_v\mathbf{e}_t+\mathbf{b}_v),$$
		where $\mathbf{v}_t$ denotes the $t$-th transformed passage signals, $\mathbf{W}_v$ and $\mathbf{b}_v$ are parameters to be learned.
We then apply a dimension-wise k-max pooling layer to select top-k signals, and feed the selected signals into a multi-layer perceptron for final assessment.

\subsection{Model Training}
Since the ad-hoc retrieval task is fundamentally a ranking problem, we utilize the pairwise ranking loss such as hinge loss to train our model. Specifically, given a triple $(q, d^+, d^-)$, where $d^+$ is ranked higher than $d^-$ with respect to a query $q$, the loss function is defined as:
$$\mathcal{L}(q, d^+, d^-; \theta) = \max(0, 1-s(q, d^+) + s(q,d^-)),$$
where $s(q,d)$ denotes the relevance score for $(q,d)$, and $\theta$ includes the parameters in both local matching layer and global decision layer. The optimization is relatively straightforward with standard backpropagation. We apply stochastic gradient decent method Adam \cite{kingma2014adam} with mini-batches(100 in size), which can be easily parallelized on a single machine with multi-cores.

\section{Experiment}
In this section, we conduct experiments to demonstrate the effectiveness of our proposed model on benchmark collections.

\begin{table}[bpt]
\centering
\caption{Statistics of the datasets used in this study.}
\begin{tabular}{c c c c c c}
\hline
 & \#queries & \#docs & \#q\_rel &  \#rel\_per\_q\\ \hline
 MQ2007 & \num{1692} & \num{65323} & \num{1455} & 10.3 \\
 MQ2008 & \num{784} & \num{14384} & \num{564} & 3.7 \\ \hline
\end{tabular}
\label{tab:data_sets}
\end{table}

\subsection{Experimental Settings}
We first introduce our experimental settings, including datasets, baseline methods/implementations, and evaluation methodology.

\subsubsection{Data Sets}
To evaluate the performance of our model, we conducted experiments on two LETOR benchmark datasets \cite{qin2010letor}: Million Query Track 2007 (MQ2007) and Million Query Track 2008 (MQ2008). We choose these two datasets according to three criteria: 1) there is a large number of queries, 2) the original document content is available, and 3) the dataset is public. The first two criterias are important for learning deep neural models for ad-hoc retrieval, and the last one is critical for reproducibility. 
Both datasets use the GOV2 collection which includes 25 million documents in 426 gigabytes.
The details of the two datasets are given in Table \ref{tab:data_sets}. As we can see, there are \num{1692} queries on MQ2007 and \num{784} queries on MQ2008. The number of queries with at least one relevant document is \num{1455} and \num{564}, respectively. The average number of relevant document per query is about $10.3$ and $3.7$ on MQ2007 and MQ2008, respectively.

For pre-processing, all the words in documents and queries are white-space tokenized, lower-cased, and stemmed using the Krovetz stemmer \cite{krovetz1993viewing}. Stopword removal is performed on query and document words using the INQUERY stop list \cite{callan1995trec}. Words occurred less than $5$ times in the collection are removed from all the document. We further segmented documents into passages for all the models using passage-level information. We utilized fixed-size sliding window without overlap to generate passages. We have also studied the performance of different window size in Section 4.5.

\begin{table*}[th]\centering
\caption{Analysis on local matching layer on MQ2007. Significant improvement or degradation with respect to our implementation ($\mathbf{S}^{\mathit{xor}}$+$\mathbf{S}^{\mathit{cos}}$+spatial GRU) is indicated (+/-) ($\text{p-value} \le 0.05 $)}
\begin{tabular}{c c c c c c c c c c }
\hline
 & \multicolumn{5}{c}{IR Heuristics} & & \multicolumn{3}{c}{Performance}\\\cline{2-6}\cline{8-10}
 Local Matching Layer & \specialcell{Exact\\matching} & \specialcell{Semantic\\ matching} & \specialcell{Exact/Semantic\\Distinguished} & Proximity & \specialcell{Term\\Importance} & & P@10 & NDCG@10 & MAP\\
\hline
$\mathbf{M}^{\mathit{xor}}$+MLP & $\surd$ &&&&&& $0.384^-$& $0.435^-$ & $0.461^-$ \\
$\mathbf{M}^{\mathit{cos}}$+MLP & $\surd$ & $\surd$ &&&&& $0.329^-$& $0.344^-$& $0.386^-$\\
$\mathbf{M}^{\mathit{hist}}$+MLP & $\surd$ & $\surd$ &$\surd$ &&&& $0.393^-$ & $0.447^-$ & $0.469^-$\\
$\mathbf{M}^{\mathit{xor}}$+spatial GRU & $\surd$ & & &$\surd$ &&& $0.387^-$& $0.444^-$ & $0.465^-$ \\
$\mathbf{M}^{\mathit{cos}}$+spatial GRU & $\surd$ & $\surd$ && $\surd$&&&$0.396^-$ &$0.449^-$& $0.470^-$ \\
$\mathbf{M}^{\mathit{xor}}$+$\mathbf{M}^{\mathit{cos}}$+spatial GRU & $\surd$ & $\surd$ & $\surd$ & $\surd$ &&&$0.405^-$ &$0.470^-$& $0.484^-$\\
$\mathbf{S}^{\mathit{xor}}$+$\mathbf{S}^{\mathit{cos}}$+spatial GRU & $\surd$ & $\surd$ & $\surd$ & $\surd$ & $\surd$ & &$0.418~~$ &$0.490~~$& $0.502~~$\\
\hline\hline
\end{tabular}
\label{tab:evidence_results}
\end{table*}

\subsubsection{Baselines Methods}
We adopt three types of baselines for comparison, including traditional retrieve models, learning to rank models and deep matching models.

For traditional retrieval models, we consider both document-wide methods, passage-level methods, and hybrid methods:
\begin{description}
	\item \textbf{BM25}: The BM25 model \cite{robertson1994some} is a classical and highly effective document-wide retrieval model.
	\item \textbf{MSP}: The Max-Scoring Passage model \cite{Liu2002Passage} utilizes language model for each passage and rank the document according to the score of their best passage.
	\item \textbf{PLM}: The passage language model \cite{bendersky2008utilizing} integrates passage-level and document-wide language model scores according to the document homogeneity for ad-hoc retrieval.
	\item \textbf{PPM}: The probabilistic passage model \cite{wang2008discriminative} is  a discriminative probabilistic model in capturing passage-level signals, and combines document retrieval scores with passage retrieval scores through a linear interpolate function.
\end{description}
Learning to rank models include
\begin{description}
	\item \textbf{AdaRank}: AdaRank \cite{xu2007adarank} is a representative pairwise model which aims to directly optimize the performance measure based on boosting approach. Here we utilize NDCG as the performance measure function.
	\item \textbf{LambdaMart}: LambdaMart \cite{burges2010ranknet} is a representative listwise model that uses gradient boosting to produce an ensemble of retrieval models. It is the state-of-the-art learning to rank algorithm.
\end{description}
Here, AdaRank and LambdaMart were implemented using RankLib\footnote{https://sourceforge.net/p/lemur/wiki/RankLib/}, which is a widely adopted learning to rank tool. All the learning to rank models leveraged the $46$ human designed features from LETOR. Furthermore, since our model utilized passage-level information, we introduced $9$ passage-based features for fair comparison. Specifically, we calculated tf-idf, BM25 and language model scores for each query-passage pair, and picked the maximum, minimum and average scores across passages as the new features for a document. We applied the full set of features (original+passage features) on both two learning to rank models for additional comparison, denoted by \textbf{AdaRank(+P)} and \textbf{LambdaMart(+P)}, respectively.

Deep matching models include
\begin{description}
	\item \textbf{DSSM}: DSSM \cite{huang2013learning} is a neural matching model proposed for Web search. It consists of a word hashing layer, two non-linear hidden layers, and an output layer.
	\item \textbf{DRMM}: DRMM \cite{guo2016deep} is a neural relevance model designed for ad-hoc retrieval. It consists of a matching histogram mapping, a feed forward matching network and a term gating network.
	\item \textbf{Duet}: Duet \cite{mitra2017learning} is a joint model which learns  local lexical matching and global semantic matching together.
	\item \textbf{DeepRank}: DeepRank \cite{pang2017deeprank} is a state-of-the-art deep matching model which models relevance by simulating the human judgement process.
\end{description}
Here, for DSSM, we directly utilize the trained model\footnote{https://www.microsoft.com/en-us/research/project/dssm/} released by their authors since training these complex models on small benchmark datasets could lead to severe over-fitting problem. 
For DeepRank\footnote{https://github.com/pl8787/textnet-release}, we use the code released by their authors. 
For Duet model,  we train it by ourselves since there is no trained model released. To avoid overfitting, we reduce the parameters of the convolutional network and fully-connected network to adapt the model to the limited size of LETOR datasets. Specifically, we set the filter size as $10$ in both local model and global model, and the hidden size as $20$ in the fully-connected layer. Other parameters are the same as the original paper. 

We refer to our proposed model as \textbf{HiNT}\footnote{https://github.com/faneshion/HiNT}. For network configurations (e.g., numbers of layers and hidden nodes), we tuned the hyper-parameters via the validation set.
Specifically, in the local matching layer,
the dimension of the spatial GRU is set to $2$ which is tuned in [1, 2, 3, 4]. In the global decision layer, the dimension of LSTM is set to $6$ which is tuned in [4, 5, 6, 7, 8, 9, 10], the k-max pooling size is set to $10$ which is tuned in [1, 5, 10, 15, 20], and the multi-layer perceptron is a 2-layers feed forward network without hidden layers.
 All the trainable parameters are initialized randomly by uniform distribution within $[-0.1,0.1]$. Overall, the number of trainable parameters is about $930$ in our HiNG model. Note that the MQ2008 dataset has much smaller query and document size, we find it is not sufficient to train deep models purely based on this dataset. Therefore, for all the deep models, we chose to use the trained model on MQ2007 as the initialization and fine tuned the model on MQ2008.

For all deep models based on term vector inputs, we used $50$-dimension term vectors. The term vectors were trained on wikipedia corpus\footnote{http://en.wikipedia.org/wiki/Wikipedia\_database} using the CBOW model \cite{mikolov2013distributed} with the default parameters\footnote{https://code.google.com/archive/p/word2vec/}. Specifically, we used $10$ as the context window size, $10$ negative samples and a subsampling of frequent words with sampling threshold of $10^{-4}$.
Out-of-vocabulary words are randomly initialized by sampling values uniformly from $(-0.02, 0.02)$.

\subsubsection{Evaluation Methodology}
We follow the data partition on this dataset in Letor4.0 \cite{qin2010letor}, and 5 fold cross-validation is conducted to minimize over-fitting as in \cite{guo2016deep}. Specifically,
the parameters for each model are tuned on 4-of-5 folds. The last fold in each case is used for evaluation. This process is repeated $5$ times, once for each fold. The results reported are the average over the $5$ folds.

As for evaluation measure, precision (P), mean average precision (MAP) and normalized discounted cumulative gain (NDCG) at position 1, 5, and 10 were used in our experiments. We performed significant tests using the paired t-test. Differences are considered statistically significant when the $p-$value is lower than $0.05$.

\subsection{Analysis on the HiNT Model}
In this section we conducted experiments to compare different implementations of the two components in the HiNT model. Through these experiments, we try to gain a better understanding of the model.

\subsubsection{Analysis on Local Matching Layer}
As mentioned in Section 3.1, the local matching layer should be able to encode many well-known IR heuristics in order to well capture the relevance matching between a query and a passage.
The heuristics at least include the modeling of exact matching and semantic matching signals, the differentiation between them, the modeling of proximity, the term importance, and so on. Here we conduct experiments to test a variety of implementations of the local matching layer which encode different IR heuristics by fixing the rest parts of the model.

The implementations include: (1) We apply a multi-layer perceptron over the exact matching matrix $\mathbf{M}^{\mathit{xor}}$ to produce the passage-level signals. In this way, only exact matching signals are encoded into the passage-level signal; (2) We apply a multi-layer perceptron over the semantic matching matrix $\mathbf{M}^{\mathit{cos}}$ to produce the passage-level signals. In this way, both exact and semantic matching signals are encoded but mixed together; (3) We follow the idea in \cite{guo2016deep} to turn the semantic matching matrix $\mathbf{M}^{\mathit{cos}}$ into matching histograms $\mathbf{M}^{\mathit{hist}}$, and use a multi-layer perceptron to produce the passage-level signals. In this way, both exact matching and semantic matching signals are encoded and these two types of signals are differentiated by using the histogram; (4) We apply a spatial GRU over the exact matching matrix $\mathbf{M}^{\mathit{xor}}$ to produce the passage-level signals. In this way, only exact matching signals and proximity are encoded into the signals; (5) We apply a spatial GRU over the semantic matching matrix $\mathbf{M}^{\mathit{cos}}$ to produce the passage-level signals. In this way, exact matching and semantic matching signals are mixed and encoded together with proximity information. (6) We use a spatial GRU over both exact matching  and semantic matching signals. Here, the exact matching and semantic matching signals can be clearly differentiated. Finally, we use our proposed implementation, i.e., a spatial GRU over the $\mathbf{S}^{\mathit{xor}}$ and $\mathbf{S}^{\mathit{cos}}$ tensors, which can encode exact matching signals, semantic matching signals, proximity and term importance. The different implementations of the local matching layer as well as their performance results are shown in Table \ref{tab:evidence_results}.

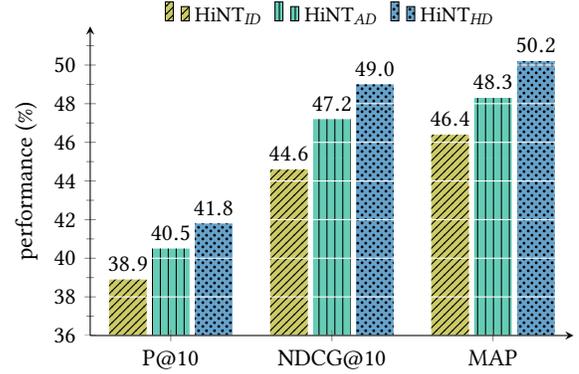
\begin{figure}[tb]
\centering	
\begin{tikzpicture}
	\begin{axis}[
		ybar,
		axis on top,
        height=.32\textwidth,
        width=.45\textwidth,
        bar width=0.5cm,
        ytick={36,38,40,42,44,46,48,50},
        ymajorgrids, tick align=inside,
        major grid style={draw=white},
        minor y tick num={1},
        enlarge y limits={value=.1,upper},
        ymin=36, ymax=52,
        axis lines=left,
        enlarge x limits=0.25,
        legend style={
            at={(0.5,1.1)},
            font=\small,
            anchor=north,
            draw=none,
            legend columns=-1,
            /tikz/every even column/.append style={column sep=0.1cm}
        },
        ylabel={performance (\%)},
        symbolic x coords={
           P@10,NDCG@10,MAP},
       xtick=data,
       nodes near coords={
        \pgfmathprintnumber[fixed zerofill,precision=1]{\pgfplotspointmeta}
       }
	]
	\addplot [draw=none, fill=green1, postaction={pattern=north east lines}] coordinates {
      (P@10,38.9)
      (NDCG@10, 44.6)
      (MAP,46.4)
      };
   \addplot [draw=none,fill=blue1, postaction={pattern=vertical lines}] coordinates {
      (P@10,40.5)
      (NDCG@10, 47.2)
      (MAP,48.3)
      };
   \addplot [draw=none, fill=red1, postaction={pattern= crosshatch dots}] coordinates {
      (P@10,41.8)
      (NDCG@10, 49.0)
      (MAP,50.2)
      };
      \legend{HiNT$_{\mathit{ID}}$,HiNT$_{\mathit{AD}}$,HiNT$_{\mathit{HD}}$}
	\end{axis}

\end{tikzpicture}
\caption{Performance comparison of HiNT over different decision models on MQ2007.}
\label{fig:decision_results}
\end{figure}

From the results we observe that, when modeling exact matching signals alone, $\mathbf{M}^{\text{xor}}+MLP$ can already obtain reasonably good retrieval performance. It indicates that exact matching signals are critical in ad-hoc retrieval \cite{guo2016deep}. Meanwhile, the performance drops significantly when semantic matching signals are mixed with exact matching signals ($\mathbf{M}^{\text{cos}}+MLP$), but increases if these two types of signals are clearly differentiated ($\mathbf{M}^{\text{hist}}+MLP$). These results demonstrate that semantic matching signals are also useful for retrieval, but should better be distinguished from exact matching signals if the deep model itself (e.g., MLP) cannot differentiate them. If the deep model can somehow implicitly distinguish these two types of signals, e.g., spatial GRU using input gates, we can observe better performance on the semantic matching matrix ($\mathbf{M}^{cos}+spatial GRU$) than that on the exact matching matrix ($\mathbf{M}^{xor}+spatial GRU$). However, we can further observe performance increase if we explicitly distinguish exact matching and semantic matching signals ($\mathbf{M}^{\text{xor}}+\mathbf{M}^{\text{cos}}+spatial GRU$).
Besides, the local matching layers using spatial GRU can in general obtain better results, indicating that proximity is very helpful for retrieval.
Finally, by further considering term importance, our proposed implementation ($\mathbf{S}^{\text{xor}}+\mathbf{S}^{\text{cos}}+spatial GRU$) can outperform all the variants significantly.
All the results demonstrate the importance of encoding a variety of IR heuristics in the local matching layer for a successful relevance judgement model.

\begin{table*}[!ht]\centering
\caption{Comparison of different retrieval models over the MQ2007 and MQ2008 datasets. Significant improvement or degradation with respect to HiNT is indicated (+/-) ($\text{p-value} \le 0.05 $).}
\begin{tabular}{c c c c c c c c c}

\multicolumn{9}{c}{MQ2007}\\
\hline\hline
Model Name & P@1 & P@5 & P@10 & NDCG@1 & NDCG@5 & NDCG@10 & MAP\\
\hline
BM25 & $0.427^-$ & $0.388^-$& $0.366^-$& $0.358^-$& $0.384^-$& $0.414^-$& $0.450^-$& \\
MSP  & $0.361^-$& $0.358^-$ & $0.350^-$ & $0.302^-$ & $0.341^-$ & $0.378^-$ & $0.422^-$\\
 PLM & $0.416^-$& $0.389^-$& $0.371^-$& $0.348^-$& $0.377^-$& $0.413^-$& $0.449^-$& \\
 PPM & $0.431^-$& $0.393^-$& $0.370^-$& $0.361^-$& $ 0.392^-$& $0.424^-$& $0.453^-$\\
 \hline
 AdaRank & $0.449^-$ & $0.403^-$& $0.372^-$& $0.394^-$ & $0.410^-$& $0.436^-$& $0.460^-$\\
  LambdaMart & $0.481^-$ & $0.418^-$& $0.384^-$& $0.412^-$ & $0.421^-$& $0.446^-$& $0.468^-$\\
  AdaRank(+P) & $0.457^-$ & $0.408^-$& $0.380^-$& $0.393^-$ & $0.408^-$& $0.438^-$& $0.467^-$\\
  LambdaMart(+P) & $0.484^-$ & $0.427^-$& $0.391^-$& $0.413^-$ & $0.427^-$& $0.454^-$& $0.473^-$\\
 \hline
 DSSM & $0.345^-$& $0.359^-$& $0.352^-$& $0.290^-$& $0.335^-$& $0.371^-$ & $0.409^-$\\
DRMM & $0.450^-$& $0.417^-$& $0.388^-$& $0.380^-$& $0.408^-$& $0.440^-$& $0.467^-$\\
 Duet & $0.473^-$& $0.428^-$& $0.398^-$& $0.409^-$& $0.431^-$& $0.453^-$& $0.474^-$\\
 DeepRank & $0.508~~$ & $0.452^-$ & $0.412^-$ & $0.441~~$ & $0.457^-$ & $0.482^-$ & $0.497~~$\\
\hline
 HiNT & \textbf{0.515~~} & \textbf{0.461~~} & \textbf{0.418~~} & \textbf{0.447~~} & \textbf{0.463~~} & \textbf{0.490~~} & \textbf{0.502~~}\\

 \hline\hline
 \multicolumn{9}{c}{MQ2008}\\
\hline\hline
Model Name & P@1 & P@5 & P@10 & NDCG@1 & NDCG@5 & NDCG@10 & MAP\\
\hline
BM25 & $0.408^-$& $0.337^-$& 0.245~~& $0.344^-$& $0.461^-$& $0.220^-$& $0.465^-$& \\
MSP & $0.332^-$& $0.314^-$& $0.236^-$& $0.283^-$& $0.415^-$& $0.193^-$& $0.426^-$\\
 PLM & $0.396^-$& $0.326^-$& $0.240^-$& $0.327^-$& $0.438^-$& $0.208^-$& $0.452^-$& \\
 PPM & $0.412^-$& $0.338^-$& $0.241^-$& $0.350^-$ & $0.464^-$& $0.220^-$ & $0.468^-$ \\
 \hline
AdaRank& $0.434^-$& 0.342~~ & 0.243~~ & $0.368^-$& 0.468~~ & 0.221~~ & 0.476~~ &\\
 LambdaMart& $0.449^-$& 0.346~~ & 0.249~~ & $0.376^-$& 0.471~~ & 0.230~~ & 0.478~~\\
 AdaRank(+P)& $0.428^-$& 0.345~~ & 0.247~~ & $0.368^-$& 0.475~~ & 0.225~~ & 0.478~~ &\\
  LambdaMart(+P)& $0.441^-$& 0.348~~& 0.249~~ & $0.372^-$& 0.479~~ & 0.232~~ & 0.480~~ \\
 \hline
 DSSM & $0.341^-$& $0.284^-$& $0.221^-$& $0.286^-$& $0.378^-$& $0.178^-$& $0.391^-$&\\
 DRMM& $0.450^-$& $0.337^-$& $0.242^-$& $0.381^-$& $0.466^-$& $0.219^-$& $0.473^-$& \\
 Duet & $0.452^-$& $0.341^-$& $0.240^-$& $0.385^-$& $0.471^-$&$0.216$& $0.476^-$ \\
 DeepRank & $0.482^-$ & $0.359^-$ & $0.252~~$ & $0.406^-$ & $0.496~~$ & $0.240~~$ & $0.498^-$\\
 \hline
 HiNT & \textbf{0.491~~} & \textbf{0.367~~} & \textbf{0.255~~} & \textbf{0.415~~} & \textbf{0.501~~} & \textbf{0.244~~} & \textbf{0.505~~} &\\
 \hline\hline
\end{tabular}
\label{tab:main_results}
\end{table*}

\subsubsection{Analysis on Global Decision Layer}
We further study the effect of different implementations of the global decision layer. Here we compare the proposed hybrid decision model (i.e., HiNT$_{\mathit{HD}}$) with the two basic decision models introduced in Section 3.2 (i.e., HiNT$_{\mathit{ID}}$ and HiNT$_{\mathit{AD}}$) by fixing the rest parts of the model.
The comparison results are shown in Figure \ref{fig:decision_results}. As we can see, the simplest relevance model HiNT$_{\mathit{ID}}$ performs worst. It seems that selecting passage-level signals independently might be too simple to capture diverse relevance patterns. Meanwhile, HiNT$_{\mathit{AD}}$ performs better than HiNT$_{\mathit{ID}}$, indicating that it is more beneficial to make relevance assessment based on accumulated signals from a variety of text spans.
Finally, HiNT$_{\mathit{HD}}$ achieves the best performance in terms of all the evaluation measures. This further indicates that there might be very diverse relevance patterns across different query-document pairs.
By allowing competition between passage-level and accumulated signals, the expressive power of the HD model is the largest, leading to the best performance among the three variants.

\begin{figure*}[tbp]
\centering
\includegraphics[scale=0.43]{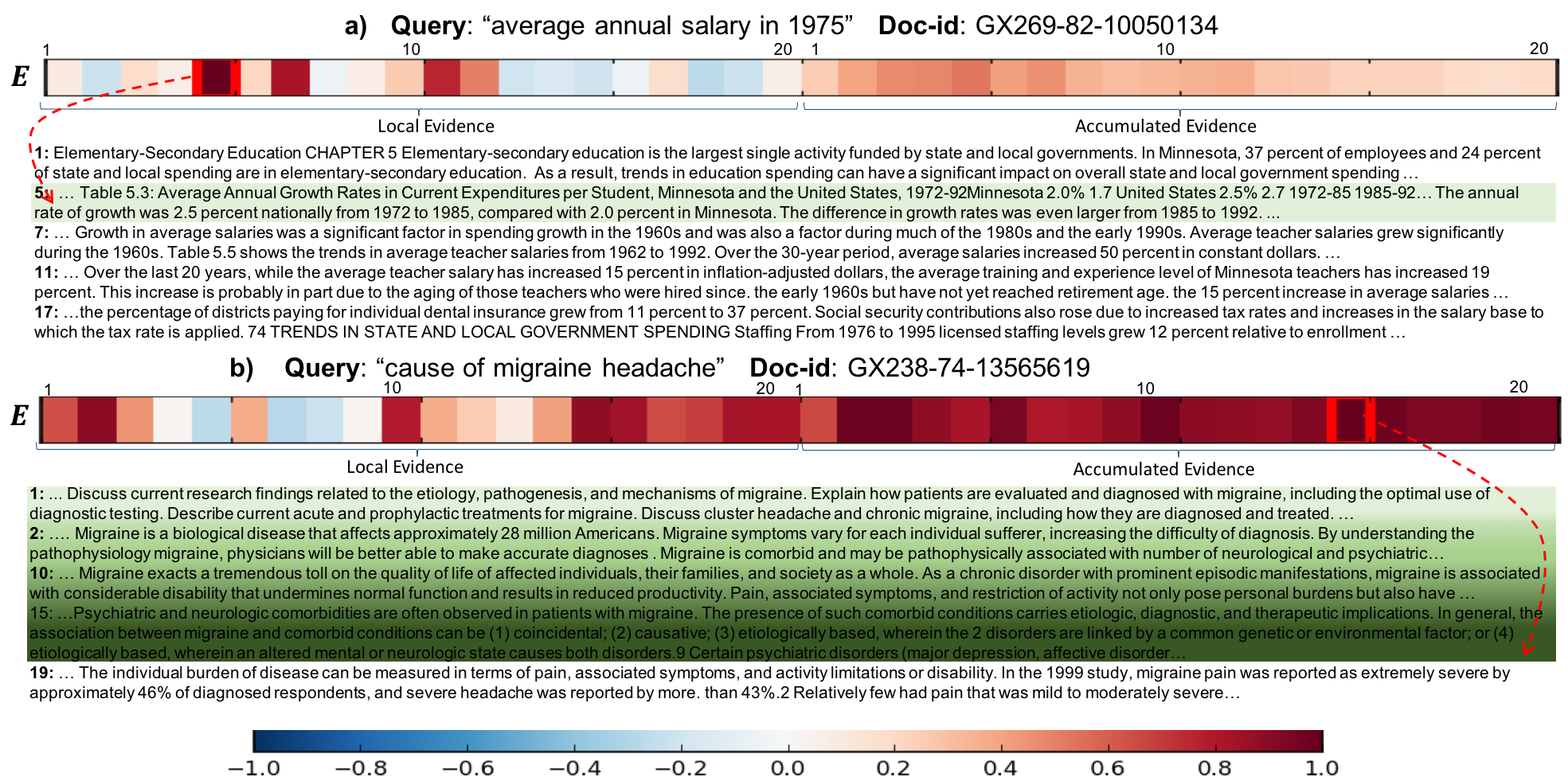}
\caption{Examples of different decision strategies over the query-document pair.}
\label{fig:case_study}
\end{figure*}

\subsection{Comparison of Retrieval Models}
In this section, we compare our proposed HiNT against existing retrieval models over the two benchmark datasets. Note here we refer HiNT to the model using hybrid decision model based on both exact matching and semantic matching tensors. The main results of our experiments are summarized in Table \ref{tab:main_results}.

Firstly, for the traditional models, we can see that BM25 is a strong baseline which performs better than MSP. The relative poor performance of MSP indicates that it is deficient in capturing the diverse relevance patterns by only using the passage-level signals.
By integrating document-wide with passage-level signals, the performance of PLM and PPM is mixed compared with BM25, demonstrating the deficiency of the simple combination strategy, which is consistent with previous findings~\cite{wang2008discriminative}.
 Secondly, all the learning to rank models perform significantly better than the traditional retrieval models. This is not surprising since learning to rank models can make use of rich features, where BM25 scores and LM scores are typical features among them. Among the two learning to rank models, LambdaMart performs better. Moreover, we can see that adding passage-level features might improve the performance of learning to rank models, but not consistent on different datasets. For example, the performance of AdaRank and LambdaMart in terms of P@1 on MQ2008 drops when adding passage features.
 Thirdly, as for the deep matching models, we can see that DSSM obtain relatively poor performances on both datasets, even cannot compete with the traditional retrieval models.
 This is consistent with previous findings \cite{guo2016deep}, showing that one may not work well on ad-hoc retrieval by only leveraging the cosine similarity between high-level abstract representations of short query and long document. 
 As for DRMM and Duet, they have achieved relative better performance compared with DSSM. This may due to the fact that they are specifically designed for the relevance matching in ad-hoc retrieval, and they have incorporated important IR heuristics into their model. However, they can only reach comparable performance as learning to rank models by only using document-wide matching signals. 
 The recently introduced DeepRank achieves a relative better performance by simulating the human judgement process.
 However, DeepRank aggregates relevance signals from query-centric contexts to form document-wide relevance score for each query term, it cannot well capture the diverse relevance patterns between query-document pairs.
 
Finally, we observe that HiNT can outperform all the existing models in terms of all the evaluation measures on both datasets by allowing the competition between the passage-level signals and document-wide signals explicitly. 
It is worth noting that in the learning to rank methods, there are many human designed features including those document-wide and passage-level matching signals, as well as those about the quality of the document (e.g., PageRank). While in our HiNT, all the assessment are purely learned from the primitive word features of queries and documents. Therefore, the superior performance of HiNT suggests the importance and effectiveness of modeling the diverse relevance patterns for ad-hoc retrieval.

\subsection{Impact of Passage Size}
Since we leverage the fixed-size sliding windows to segment a document into passages, we would like to study the effect of the passage size on the ranking performance. Here we report the performance results on MQ2007 with the passage size set as $10$, $50$, $100$, $200$, $300$ and $500$ words. As shown in Figure \ref{fig:window_size}, the performance first increases and then drops with the increase of the passage size. The possible reason might be that too small passage size may hurt the quality of passage signals (i.e., relevance matching between the query and the passage) due to the information sparsity, while too large passage size would produce limited number of coarse passage-level signals which restrict the ability of the global decision layer. Our results shows that the best performance can be achieved when the passage size is set to $100$ words on MQ2007.

\pgfplotsset{
axis background/.style={fill=white},
grid=both,
  xtick pos=left,
  ytick pos=left,
  tick style={
    major grid style={style=gallery,line width=1pt},
    minor grid style=mercury,
    },
  minor tick num=1,
}

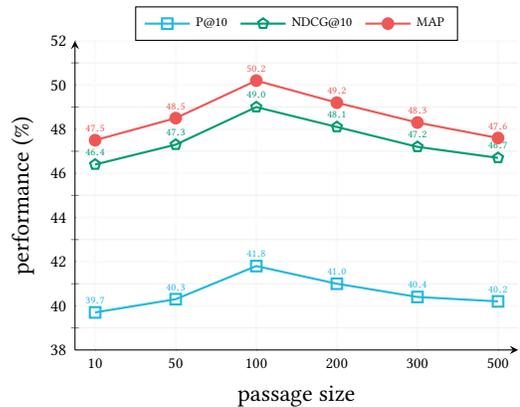
\begin{figure}[!tb]
\centering
  \begin{tikzpicture}
    \begin{axis}[
      height=.32\textwidth,
      width=.42\textwidth,
      xticklabels={10,50,100,200,300,500},
      xtick={1,2,3,4,5,6},
      legend style={
          font=\tiny,
          legend columns=-1,
          at={(0.5,1)},
          anchor=south,
          /tikz/every even column/.append style={column sep=0.9mm}
        },
        ymajorgrids={true},
        ymin=38,
        ymax=52,
        ytick={38, 40, ..., 52},
        minor x tick num={0},
        minor y tick num={1},
        axis lines=left,
        enlarge x limits=0.05,
        tickwidth=0pt,
        nodes near coords,
        every node near coord/.append style={anchor=north, font=\fontsize{4pt}{4pt}\selectfont},
        legend entries = {P@10, NDCG@10, MAP},
        xlabel={passage size},
        ylabel={performance (\%)},
        every tick label/.append style={font=\scriptsize},
        nodes near coords={
        \pgfmathprintnumber[fixed zerofill,precision=1]{\pgfplotspointmeta}
       	}
        ]

      \addplot[color=shakespeare,mark=square, every node near coord/.append style={anchor=south}, thick] coordinates {
        (1, 39.7)
        (2, 40.3)
        (3, 41.8)
        (4, 41.0)
        (5, 40.4)
        (6, 40.2)
      };
      \addplot[color=free_speech_aquamarine, mark=pentagon, every node near coord/.append style={anchor=south}, thick] coordinates {
        (1, 46.4)
        (2, 47.3)
        (3, 49.0)
        (4, 48.1)
        (5, 47.2)
        (6, 46.7)
      };
      \addplot[color=flamingo, mark=*, every node near coord/.append style={anchor=south},thick] coordinates {
        (1, 47.5)
        (2, 48.5)
        (3, 50.2)
        (4, 49.2)
        (5, 48.3)
        (6, 47.6)
      };

    \end{axis}
  \end{tikzpicture}
  \caption{Performance comparison of HiNT over different passage sizes on MQ2007.}
  \label{fig:window_size}
\end{figure}

\subsection{Case Study}
To better understand what can be learned by HiNT, here we conduct some case studies. For better visualization and analysis, we simplified our model by replacing k-max pooling with max pooling so that only the most significant signal is used in decision. Based on the learned model, we pick up a query and a relevant document, and plot all the signals $\mathbf{E}$ used in the hybrid decision model along with the corresponding document content. Here each small bar in $\mathbf{E}$ denotes a passage or accumulated signal at that position, with the color corresponding to the signal strength. We highlight the final selected signal with a red box and the corresponding passages with green background color.

As shown in Figure \ref{fig:case_study}, we can find two significantly different decision strategies between a query and a document. In the first case, the document is relevant to a query because of a strong passage signal. By checking the query and the document, we find that the query is ``average annual salary in 1975'' which conveys very specific information need, and the passage at the $5$-th position (i.e., the strongest signal) contains a table whose content can well address this information need. In the second case, the document is relevant to a query because of a strong accumulated signal. Again by checking the query and the document, we find that the query is about the ``cause of migraine headache'' which is informational, and the document is mostly relevant to this query with many passages addressing this problem (i.e., from the beginning to the $15$-th passage).

The two cases show that there are indeed quite diverse relevance patterns in real-world retrieval scenario, and our HiNT model can capture these diverse relevance patterns successfully. 

\section{Conclusions}
In this paper, we have introduced a hierarchical neural matching model to capture the diverse relevance patterns in ad-hoc retrieval. The model consists of two components, namely local matching layer and global decision layer. We employed deep neural network in both layers to support high-quality relevance signal generation and flexible relevance assessment strategies, respectively.
Experimental results on two benchmark datasets demonstrate that our model can outperform all the baseline models in terms to all the evaluation metrics, especially state-of-the-art learning to rank methods that use manually designed features.

For future work, it would be interesting to
try other implementation of the two components in HiNT, e.g., to employ some attention-based neural network for the global decision layer.
We would also like to expand our model to accommodate features beyond relevance matching, e.g. PageRank, to help improve the retrieval performance.

\section{Acknowledgments}
This work was funded by the 973 Program of China under Grant No.~2014CB340401, the National Natural Science Foundation of China (NSFC) under Grants No.~61425016, 61472401, 61722211, and 20180290, the Youth Innovation Promotion Association CAS under Grants No.~20144310 and 2016102, and the National Key R\&D Program of China under Grants No.~2016QY02D0405. 

\bibliographystyle{ACM-Reference-Format}

\begin{thebibliography}{00}


\ifx \showCODEN    \undefined \def \showCODEN     #1{\unskip}     \fi
\ifx \showDOI      \undefined \def \showDOI       #1{{\tt DOI:}\penalty0{#1}\ }
  \fi
\ifx \showISBNx    \undefined \def \showISBNx     #1{\unskip}     \fi
\ifx \showISBNxiii \undefined \def \showISBNxiii  #1{\unskip}     \fi
\ifx \showISSN     \undefined \def \showISSN      #1{\unskip}     \fi
\ifx \showLCCN     \undefined \def \showLCCN      #1{\unskip}     \fi
\ifx \shownote     \undefined \def \shownote      #1{#1}          \fi
\ifx \showarticletitle \undefined \def \showarticletitle #1{#1}   \fi
\ifx \showURL      \undefined \def \showURL       #1{#1}          \fi
\providecommand\bibfield[2]{#2}
\providecommand\bibinfo[2]{#2}
\providecommand\natexlab[1]{#1}
\providecommand\showeprint[2][]{arXiv:#2}

\bibitem[\protect\citeauthoryear{Amati and Van~Rijsbergen}{Amati and
  Van~Rijsbergen}{2002}]%
        {amati2002probabilistic}
\bibfield{author}{\bibinfo{person}{Gianni Amati} {and}
  \bibinfo{person}{Cornelis~Joost Van~Rijsbergen}.}
  \bibinfo{year}{2002}\natexlab{}.
\newblock \showarticletitle{Probabilistic models of information retrieval based
  on measuring the divergence from randomness}.
\newblock \bibinfo{journal}{{\em ACM Transactions on Information Systems
  (TOIS)\/}} \bibinfo{volume}{20}, \bibinfo{number}{4} (\bibinfo{year}{2002}),
  \bibinfo{pages}{357--389}.
\newblock


\bibitem[\protect\citeauthoryear{Bendersky and Kurland}{Bendersky and
  Kurland}{2008}]%
        {bendersky2008utilizing}
\bibfield{author}{\bibinfo{person}{Michael Bendersky} {and}
  \bibinfo{person}{Oren Kurland}.} \bibinfo{year}{2008}\natexlab{}.
\newblock \showarticletitle{Utilizing passage-based language models for
  document retrieval}. In \bibinfo{booktitle}{{\em European Conference on
  Information Retrieval}}. Springer, \bibinfo{pages}{162--174}.
\newblock


\bibitem[\protect\citeauthoryear{Burges}{Burges}{2010}]%
        {burges2010ranknet}
\bibfield{author}{\bibinfo{person}{Christopher~JC Burges}.}
  \bibinfo{year}{2010}\natexlab{}.
\newblock \showarticletitle{From ranknet to lambdarank to lambdamart: An
  overview}.
\newblock \bibinfo{journal}{{\em Learning\/}}  \bibinfo{volume}{11}
  (\bibinfo{year}{2010}), \bibinfo{pages}{23--581}.
\newblock


\bibitem[\protect\citeauthoryear{Callan}{Callan}{1994}]%
        {callan1994passage}
\bibfield{author}{\bibinfo{person}{James~P Callan}.}
  \bibinfo{year}{1994}\natexlab{}.
\newblock \showarticletitle{Passage-level evidence in document retrieval}. In
  \bibinfo{booktitle}{{\em SIGIR}}. Springer-Verlag New York, Inc.,
  \bibinfo{pages}{302--310}.
\newblock


\bibitem[\protect\citeauthoryear{Callan, Croft, and Broglio}{Callan
  et~al\mbox{.}}{1995}]%
        {callan1995trec}
\bibfield{author}{\bibinfo{person}{James~P Callan}, \bibinfo{person}{W~Bruce
  Croft}, {and} \bibinfo{person}{John Broglio}.}
  \bibinfo{year}{1995}\natexlab{}.
\newblock \showarticletitle{TREC and TIPSTER experiments with INQUERY}.
\newblock \bibinfo{journal}{{\em Information Processing \& Management\/}}
  \bibinfo{volume}{31}, \bibinfo{number}{3} (\bibinfo{year}{1995}),
  \bibinfo{pages}{327--343}.
\newblock


\bibitem[\protect\citeauthoryear{Clarke, Scholer, and Soboroff}{Clarke
  et~al\mbox{.}}{2005}]%
        {clarke2005trec}
\bibfield{author}{\bibinfo{person}{Charles~LA Clarke}, \bibinfo{person}{Falk
  Scholer}, {and} \bibinfo{person}{Ian Soboroff}.}
  \bibinfo{year}{2005}\natexlab{}.
\newblock \showarticletitle{The TREC 2005 Terabyte Track.}. In
  \bibinfo{booktitle}{{\em TREC}}.
\newblock


\bibitem[\protect\citeauthoryear{Fang and Zhai}{Fang and Zhai}{2006}]%
        {fang2006semantic}
\bibfield{author}{\bibinfo{person}{Hui Fang} {and} \bibinfo{person}{ChengXiang
  Zhai}.} \bibinfo{year}{2006}\natexlab{}.
\newblock \showarticletitle{Semantic term matching in axiomatic approaches to
  information retrieval}. In \bibinfo{booktitle}{{\em SIGIR}}. ACM,
  \bibinfo{pages}{115--122}.
\newblock


\bibitem[\protect\citeauthoryear{Freund, Iyer, Schapire, and Singer}{Freund
  et~al\mbox{.}}{2003}]%
        {freund2003efficient}
\bibfield{author}{\bibinfo{person}{Yoav Freund}, \bibinfo{person}{Raj Iyer},
  \bibinfo{person}{Robert~E Schapire}, {and} \bibinfo{person}{Yoram Singer}.}
  \bibinfo{year}{2003}\natexlab{}.
\newblock \showarticletitle{An efficient boosting algorithm for combining
  preferences}.
\newblock \bibinfo{journal}{{\em Journal of machine learning research\/}}
  \bibinfo{volume}{4}, \bibinfo{number}{Nov} (\bibinfo{year}{2003}),
  \bibinfo{pages}{933--969}.
\newblock


\bibitem[\protect\citeauthoryear{Graves and Schmidhuber}{Graves and
  Schmidhuber}{2009}]%
        {graves2009offline}
\bibfield{author}{\bibinfo{person}{Alex Graves} {and}
  \bibinfo{person}{J{\"u}rgen Schmidhuber}.} \bibinfo{year}{2009}\natexlab{}.
\newblock \showarticletitle{Offline handwriting recognition with
  multidimensional recurrent neural networks}. In \bibinfo{booktitle}{{\em
  Advances in neural information processing systems}}.
  \bibinfo{pages}{545--552}.
\newblock


\bibitem[\protect\citeauthoryear{Guo, Fan, Ai, and Croft}{Guo
  et~al\mbox{.}}{2016a}]%
        {guo2016deep}
\bibfield{author}{\bibinfo{person}{Jiafeng Guo}, \bibinfo{person}{Yixing Fan},
  \bibinfo{person}{Qingyao Ai}, {and} \bibinfo{person}{W~Bruce Croft}.}
  \bibinfo{year}{2016}\natexlab{a}.
\newblock \showarticletitle{A deep relevance matching model for ad-hoc
  retrieval}. In \bibinfo{booktitle}{{\em CIKM}}. ACM, \bibinfo{pages}{55--64}.
\newblock


\bibitem[\protect\citeauthoryear{Guo, Fan, Ai, and Croft}{Guo
  et~al\mbox{.}}{2016b}]%
        {guo2016semantic}
\bibfield{author}{\bibinfo{person}{Jiafeng Guo}, \bibinfo{person}{Yixing Fan},
  \bibinfo{person}{Qingyao Ai}, {and} \bibinfo{person}{W~Bruce Croft}.}
  \bibinfo{year}{2016}\natexlab{b}.
\newblock \showarticletitle{Semantic matching by non-linear word transportation
  for information retrieval}. In \bibinfo{booktitle}{{\em CIKM}}. ACM,
  \bibinfo{pages}{701--710}.
\newblock


\bibitem[\protect\citeauthoryear{Hochreiter and Schmidhuber}{Hochreiter and
  Schmidhuber}{1997}]%
        {hochreiter1997long}
\bibfield{author}{\bibinfo{person}{Sepp Hochreiter} {and}
  \bibinfo{person}{J{\"u}rgen Schmidhuber}.} \bibinfo{year}{1997}\natexlab{}.
\newblock \showarticletitle{Long short-term memory}.
\newblock \bibinfo{journal}{{\em Neural computation\/}} \bibinfo{volume}{9},
  \bibinfo{number}{8} (\bibinfo{year}{1997}), \bibinfo{pages}{1735--1780}.
\newblock


\bibitem[\protect\citeauthoryear{Hu, Lu, Li, and Chen}{Hu
  et~al\mbox{.}}{2014}]%
        {hu2014convolutional}
\bibfield{author}{\bibinfo{person}{Baotian Hu}, \bibinfo{person}{Zhengdong Lu},
  \bibinfo{person}{Hang Li}, {and} \bibinfo{person}{Qingcai Chen}.}
  \bibinfo{year}{2014}\natexlab{}.
\newblock \showarticletitle{Convolutional neural network architectures for
  matching natural language sentences}. In \bibinfo{booktitle}{{\em NIPS}}.
  \bibinfo{pages}{2042--2050}.
\newblock


\bibitem[\protect\citeauthoryear{Huang, He, Gao, Deng, Acero, and Heck}{Huang
  et~al\mbox{.}}{2013}]%
        {huang2013learning}
\bibfield{author}{\bibinfo{person}{Po-Sen Huang}, \bibinfo{person}{Xiaodong
  He}, \bibinfo{person}{Jianfeng Gao}, \bibinfo{person}{Li Deng},
  \bibinfo{person}{Alex Acero}, {and} \bibinfo{person}{Larry Heck}.}
  \bibinfo{year}{2013}\natexlab{}.
\newblock \showarticletitle{Learning deep structured semantic models for web
  search using clickthrough data}. In \bibinfo{booktitle}{{\em CIKM}}. ACM,
  \bibinfo{pages}{2333--2338}.
\newblock


\bibitem[\protect\citeauthoryear{Joachims}{Joachims}{2006}]%
        {joachims2006training}
\bibfield{author}{\bibinfo{person}{Thorsten Joachims}.}
  \bibinfo{year}{2006}\natexlab{}.
\newblock \showarticletitle{Training linear SVMs in linear time}. In
  \bibinfo{booktitle}{{\em SIGKDD}}. ACM, \bibinfo{pages}{217--226}.
\newblock


\bibitem[\protect\citeauthoryear{Kaszkiel and Zobel}{Kaszkiel and
  Zobel}{1997}]%
        {kaszkiel1997passage}
\bibfield{author}{\bibinfo{person}{Marcin Kaszkiel} {and}
  \bibinfo{person}{Justin Zobel}.} \bibinfo{year}{1997}\natexlab{}.
\newblock \showarticletitle{Passage retrieval revisited}. In
  \bibinfo{booktitle}{{\em SIGIR}}, Vol.~\bibinfo{volume}{31}. ACM,
  \bibinfo{pages}{178--185}.
\newblock


\bibitem[\protect\citeauthoryear{Kingma and Ba}{Kingma and Ba}{2014}]%
        {kingma2014adam}
\bibfield{author}{\bibinfo{person}{Diederik Kingma} {and}
  \bibinfo{person}{Jimmy Ba}.} \bibinfo{year}{2014}\natexlab{}.
\newblock \showarticletitle{Adam: A method for stochastic optimization}.
\newblock \bibinfo{journal}{{\em arXiv preprint arXiv:1412.6980\/}}
  (\bibinfo{year}{2014}).
\newblock


\bibitem[\protect\citeauthoryear{Krovetz}{Krovetz}{1993}]%
        {krovetz1993viewing}
\bibfield{author}{\bibinfo{person}{Robert Krovetz}.}
  \bibinfo{year}{1993}\natexlab{}.
\newblock \showarticletitle{Viewing morphology as an inference process}. In
  \bibinfo{booktitle}{{\em SIGIR}}. ACM, \bibinfo{pages}{191--202}.
\newblock


\bibitem[\protect\citeauthoryear{Lee}{Lee}{1997}]%
        {lee1997analyses}
\bibfield{author}{\bibinfo{person}{Joon~Ho Lee}.}
  \bibinfo{year}{1997}\natexlab{}.
\newblock \showarticletitle{Analyses of multiple evidence combination}. In
  \bibinfo{booktitle}{{\em ACM SIGIR Forum}}, Vol.~\bibinfo{volume}{31}. ACM,
  \bibinfo{pages}{267--276}.
\newblock


\bibitem[\protect\citeauthoryear{Liu and Croft}{Liu and Croft}{2002}]%
        {Liu2002Passage}
\bibfield{author}{\bibinfo{person}{Xiaoyong Liu} {and}
  \bibinfo{person}{W.~Bruce Croft}.} \bibinfo{year}{2002}\natexlab{}.
\newblock \showarticletitle{Passage retrieval based on language models}. In
  \bibinfo{booktitle}{{\em CIKM}}. \bibinfo{pages}{375--382}.
\newblock


\bibitem[\protect\citeauthoryear{Lv and Zhai}{Lv and Zhai}{2009}]%
        {Lv2009Positional}
\bibfield{author}{\bibinfo{person}{Yuanhua Lv} {and}
  \bibinfo{person}{Cheng~Xiang Zhai}.} \bibinfo{year}{2009}\natexlab{}.
\newblock \showarticletitle{Positional language models for information
  retrieval}. In \bibinfo{booktitle}{{\em SIGIR}}. \bibinfo{pages}{299--306}.
\newblock


\bibitem[\protect\citeauthoryear{Mikolov, Sutskever, Chen, Corrado, and
  Dean}{Mikolov et~al\mbox{.}}{2013}]%
        {mikolov2013distributed}
\bibfield{author}{\bibinfo{person}{Tomas Mikolov}, \bibinfo{person}{Ilya
  Sutskever}, \bibinfo{person}{Kai Chen}, \bibinfo{person}{Greg~S Corrado},
  {and} \bibinfo{person}{Jeff Dean}.} \bibinfo{year}{2013}\natexlab{}.
\newblock \showarticletitle{Distributed representations of words and phrases
  and their compositionality}. In \bibinfo{booktitle}{{\em Advances in neural
  information processing systems}}. \bibinfo{pages}{3111--3119}.
\newblock


\bibitem[\protect\citeauthoryear{Mitra and Craswell}{Mitra and
  Craswell}{2017}]%
        {mitra2017neural}
\bibfield{author}{\bibinfo{person}{Bhaskar Mitra} {and} \bibinfo{person}{Nick
  Craswell}.} \bibinfo{year}{2017}\natexlab{}.
\newblock \showarticletitle{Neural Models for Information Retrieval}.
\newblock \bibinfo{journal}{{\em arXiv preprint arXiv:1705.01509\/}}
  (\bibinfo{year}{2017}).
\newblock


\bibitem[\protect\citeauthoryear{Mitra, Diaz, and Craswell}{Mitra
  et~al\mbox{.}}{2017}]%
        {mitra2017learning}
\bibfield{author}{\bibinfo{person}{Bhaskar Mitra}, \bibinfo{person}{Fernando
  Diaz}, {and} \bibinfo{person}{Nick Craswell}.}
  \bibinfo{year}{2017}\natexlab{}.
\newblock \showarticletitle{Learning to Match using Local and Distributed
  Representations of Text for Web Search}. In \bibinfo{booktitle}{{\em
  Proceedings of the 26th International Conference on World Wide Web}}.
  International World Wide Web Conferences Steering Committee,
  \bibinfo{pages}{1291--1299}.
\newblock


\bibitem[\protect\citeauthoryear{Na}{Na}{2015}]%
        {na2015two}
\bibfield{author}{\bibinfo{person}{Seung-Hoon Na}.}
  \bibinfo{year}{2015}\natexlab{}.
\newblock \showarticletitle{Two-stage document length normalization for
  information retrieval}.
\newblock \bibinfo{journal}{{\em ACM Transactions on Information Systems
  (TOIS)\/}} \bibinfo{volume}{33}, \bibinfo{number}{2} (\bibinfo{year}{2015}),
  \bibinfo{pages}{8}.
\newblock


\bibitem[\protect\citeauthoryear{Pang, Lan, Guo, Xu, and Cheng}{Pang
  et~al\mbox{.}}{2016}]%
        {pang2016study}
\bibfield{author}{\bibinfo{person}{Liang Pang}, \bibinfo{person}{Yanyan Lan},
  \bibinfo{person}{Jiafeng Guo}, \bibinfo{person}{Jun Xu}, {and}
  \bibinfo{person}{Xueqi Cheng}.} \bibinfo{year}{2016}\natexlab{}.
\newblock \showarticletitle{A study of matchpyramid models on ad-hoc
  retrieval}.
\newblock \bibinfo{journal}{{\em arXiv preprint arXiv:1606.04648\/}}
  (\bibinfo{year}{2016}).
\newblock


\bibitem[\protect\citeauthoryear{Pang, Lan, Guo, Xu, Xu, and Cheng}{Pang
  et~al\mbox{.}}{2017}]%
        {pang2017deeprank}
\bibfield{author}{\bibinfo{person}{Liang Pang}, \bibinfo{person}{Yanyan Lan},
  \bibinfo{person}{Jiafeng Guo}, \bibinfo{person}{Jun Xu},
  \bibinfo{person}{Jingfang Xu}, {and} \bibinfo{person}{Xueqi Cheng}.}
  \bibinfo{year}{2017}\natexlab{}.
\newblock \showarticletitle{DeepRank: A New Deep Architecture for Relevance
  Ranking in Information Retrieval}. In \bibinfo{booktitle}{{\em CIKM}}. ACM,
  \bibinfo{pages}{257--266}.
\newblock


\bibitem[\protect\citeauthoryear{Qin, Liu, Xu, and Li}{Qin
  et~al\mbox{.}}{2010}]%
        {qin2010letor}
\bibfield{author}{\bibinfo{person}{Tao Qin}, \bibinfo{person}{Tie-Yan Liu},
  \bibinfo{person}{Jun Xu}, {and} \bibinfo{person}{Hang Li}.}
  \bibinfo{year}{2010}\natexlab{}.
\newblock \showarticletitle{LETOR: A benchmark collection for research on
  learning to rank for information retrieval}.
\newblock \bibinfo{journal}{{\em Information Retrieval\/}}
  \bibinfo{volume}{13}, \bibinfo{number}{4} (\bibinfo{year}{2010}),
  \bibinfo{pages}{346--374}.
\newblock


\bibitem[\protect\citeauthoryear{Rayner}{Rayner}{1998}]%
        {rayner1998eye}
\bibfield{author}{\bibinfo{person}{Keith Rayner}.}
  \bibinfo{year}{1998}\natexlab{}.
\newblock \showarticletitle{Eye movements in reading and information
  processing: 20 years of research.}
\newblock \bibinfo{journal}{{\em Psychological bulletin\/}}
  \bibinfo{volume}{124}, \bibinfo{number}{3} (\bibinfo{year}{1998}),
  \bibinfo{pages}{372}.
\newblock


\bibitem[\protect\citeauthoryear{Robertson and Walker}{Robertson and
  Walker}{1994}]%
        {robertson1994some}
\bibfield{author}{\bibinfo{person}{Stephen~E Robertson} {and}
  \bibinfo{person}{Steve Walker}.} \bibinfo{year}{1994}\natexlab{}.
\newblock \showarticletitle{Some simple effective approximations to the
  2-poisson model for probabilistic weighted retrieval}. In
  \bibinfo{booktitle}{{\em SIGIR}}. Springer-Verlag New York, Inc.,
  \bibinfo{pages}{232--241}.
\newblock


\bibitem[\protect\citeauthoryear{Salton, Allan, and Buckley}{Salton
  et~al\mbox{.}}{1993}]%
        {salton1993approaches}
\bibfield{author}{\bibinfo{person}{Gerard Salton}, \bibinfo{person}{James
  Allan}, {and} \bibinfo{person}{Chris Buckley}.}
  \bibinfo{year}{1993}\natexlab{}.
\newblock \showarticletitle{Approaches to passage retrieval in full text
  information systems}. In \bibinfo{booktitle}{{\em SIGIR}}. ACM,
  \bibinfo{pages}{49--58}.
\newblock


\bibitem[\protect\citeauthoryear{Sanderson}{Sanderson}{2010}]%
        {sanderson2010test}
\bibfield{author}{\bibinfo{person}{Mark Sanderson}.}
  \bibinfo{year}{2010}\natexlab{}.
\newblock \bibinfo{booktitle}{{\em Test collection based evaluation of
  information retrieval systems}}.
\newblock \bibinfo{publisher}{Now Publishers Inc}.
\newblock


\bibitem[\protect\citeauthoryear{Tao and Zhai}{Tao and Zhai}{2007}]%
        {tao2007exploration}
\bibfield{author}{\bibinfo{person}{Tao Tao} {and} \bibinfo{person}{ChengXiang
  Zhai}.} \bibinfo{year}{2007}\natexlab{}.
\newblock \showarticletitle{An exploration of proximity measures in information
  retrieval}. In \bibinfo{booktitle}{{\em SIGIR}}. ACM,
  \bibinfo{pages}{295--302}.
\newblock


\bibitem[\protect\citeauthoryear{Wan, Lan, Xu, Guo, Pang, and Cheng}{Wan
  et~al\mbox{.}}{2016}]%
        {wan2016match}
\bibfield{author}{\bibinfo{person}{Shengxian Wan}, \bibinfo{person}{Yanyan
  Lan}, \bibinfo{person}{Jun Xu}, \bibinfo{person}{Jiafeng Guo},
  \bibinfo{person}{Liang Pang}, {and} \bibinfo{person}{Xueqi Cheng}.}
  \bibinfo{year}{2016}\natexlab{}.
\newblock \showarticletitle{Match-SRNN: Modeling the Recursive Matching
  Structure with Spatial RNN}.
\newblock \bibinfo{journal}{{\em arXiv preprint arXiv:1604.04378\/}}
  (\bibinfo{year}{2016}).
\newblock


\bibitem[\protect\citeauthoryear{Wang and Si}{Wang and Si}{2008}]%
        {wang2008discriminative}
\bibfield{author}{\bibinfo{person}{Mengqiu Wang} {and} \bibinfo{person}{Luo
  Si}.} \bibinfo{year}{2008}\natexlab{}.
\newblock \showarticletitle{Discriminative probabilistic models for passage
  based retrieval}. In \bibinfo{booktitle}{{\em SIGIR}}. ACM,
  \bibinfo{pages}{419--426}.
\newblock


\bibitem[\protect\citeauthoryear{Wu, Luk, Wong, and Kwok}{Wu
  et~al\mbox{.}}{2007}]%
        {wu2007retrospective}
\bibfield{author}{\bibinfo{person}{Ho~Chung Wu}, \bibinfo{person}{Robert~WP
  Luk}, \bibinfo{person}{Kam-Fai Wong}, {and} \bibinfo{person}{KL Kwok}.}
  \bibinfo{year}{2007}\natexlab{}.
\newblock \showarticletitle{A retrospective study of a hybrid document-context
  based retrieval model}.
\newblock \bibinfo{journal}{{\em Information processing \& management\/}}
  \bibinfo{volume}{43}, \bibinfo{number}{5} (\bibinfo{year}{2007}),
  \bibinfo{pages}{1308--1331}.
\newblock


\bibitem[\protect\citeauthoryear{Xi, Xu-Rong, Khoo, and Lim}{Xi
  et~al\mbox{.}}{2001}]%
        {xi2001incorporating}
\bibfield{author}{\bibinfo{person}{Wensi Xi}, \bibinfo{person}{Richard
  Xu-Rong}, \bibinfo{person}{Christopher~SG Khoo}, {and}
  \bibinfo{person}{Ee-Peng Lim}.} \bibinfo{year}{2001}\natexlab{}.
\newblock \showarticletitle{Incorporating window-based passage-level evidence
  in document retrieval}.
\newblock \bibinfo{journal}{{\em Journal of information science\/}}
  \bibinfo{volume}{27}, \bibinfo{number}{2} (\bibinfo{year}{2001}),
  \bibinfo{pages}{73--80}.
\newblock


\bibitem[\protect\citeauthoryear{Xu and Li}{Xu and Li}{2007}]%
        {xu2007adarank}
\bibfield{author}{\bibinfo{person}{Jun Xu} {and} \bibinfo{person}{Hang Li}.}
  \bibinfo{year}{2007}\natexlab{}.
\newblock \showarticletitle{Adarank: a boosting algorithm for information
  retrieval}. In \bibinfo{booktitle}{{\em SIGIR}}. ACM,
  \bibinfo{pages}{391--398}.
\newblock


\bibitem[\protect\citeauthoryear{Zhai and Lafferty}{Zhai and Lafferty}{2001}]%
        {zhai2001study}
\bibfield{author}{\bibinfo{person}{Chengxiang Zhai} {and} \bibinfo{person}{John
  Lafferty}.} \bibinfo{year}{2001}\natexlab{}.
\newblock \showarticletitle{A Study of Smoothing Methods for Language Models
  Applied to Ad Hoc Information Retrieval}. In \bibinfo{booktitle}{{\em
  SIGIR}}. \bibinfo{publisher}{ACM}, \bibinfo{address}{New York, NY, USA},
  \bibinfo{pages}{334--342}.
\newblock
\showISBNx{1-58113-331-6}
\showDOI{%
\url{http://dx.doi.org/10.1145/383952.384019}}


\end{thebibliography}

\end{document}